# Application Components Migration in NFV-based Hybrid Cloud/Fog Systems

Seyedeh Negar Afrasiabi[¥], Somayeh Kianpisheh[¥], Carla Mouradian[¥], Roch H. Glitho[¥], Ashok Moghe[€]

[¥]CIISE, Concordia University, Montreal, QC, Canada,

[€]CISCO Systems, San Francisco Bay Area, USA

[¥]{s_afra, s_kianpi, ca_moura, glitho}@encs.concordia.ca, [€]amoghe@cisco.com

*Abstract*—Fog computing extends the cloud to the edge of the network, close to the end-users enabling the deployment of some application component in the fog while others in the cloud. Network Functions Virtualization (NFV) decouples the network functions from the underlying hardware. In NFV settings, application components can be implemented as sets of Virtual Network Functions (VNFs) chained in specific order representing VNF-Forwarding Graphs (VNF-FG). Many studies have been carried out to map the VNF-FGs to cloud systems. However, in hybrid cloud/fog systems, an additional challenge arises. The mobility of fog nodes may cause high latency as the distance between the end-users and the nodes hosting the components increases. This may not be tolerable for some applications. In such cases, a prominent solution is to migrate application components to a closer fog node. This paper focuses on application component migration in NFV-based hybrid cloud/fog systems. The objective is to minimize the aggregated makespan of the applications. The problem is modeled mathematically, and a heuristic is proposed to find the sub-optimal solution in an acceptable time. The heuristic aims at finding the optimal fog node in each time-slot considering a pre-knowledge of the mobility models of the fog nodes. The experiment's results show that our proposed solution improves the makespan and the number of migrations compared to random migration and No-migration.

## I. INTRODUCTION

Cloud computing has several characteristics such as scalability, on-demand resource allocation, pay-as-you-go, and easy application and services provisioning making it a distinct paradigm. However, there still exist some challenges. The cloud providers are far from their end-users causing latency which brings trouble for latency sensitive applications. To solve this issue, the new paradigm, fog computing enables processing at the edge of the network by presenting intermediate levels between end-users and the cloud [1].

Network Functions Virtualization (NFV) is a new paradigm which decouples network functions from hardware through running the functions as software instances in virtual machines or containers [1]. Application components can be implemented as Virtualized Network Functions (VNFs) that can be chained as VNF-Forwarding Graphs (VNF-FG) [2][3].

Many studies have been carried out to embed the VNF-FG in cloud systems [4][5]. Recently, few embedding methods have been proposed in hybrid cloud/fog systems [2][3]. However, the fact that fog nodes can be mobile [5] brings new challenges such as component migration. More precisely, as a result of a fog node's mobility, the hosted component may become farther from end-users which results in high latency. In this case, migrating the component to a closer fog node helps in reducing the end to end latency. Furthermore, the recent advances in running VNFs in containers [6] reduces the migration overhead and makes migration solutions practical [7]. Migration has been extensively studied in cloud systems [8][9] and mobile edge computing [12]. However, to the best of our knowledge, no migration techniques have been provided for the case that there are interactions between the components as demanded by applications such as fire detection and autonomous driving [10] [3].

In this paper, we target the Application Components Migration (ACM) problem in the NFV-based hybrid cloud/fog systems. We model the problem as an optimization problem which minimizes the aggregated makespan of all input applications. We discretize the time and with the pre-knowledge of the mobility model of the fog nodes, the migration of components may happen at time slots. A heuristic is proposed and validated by simulations. Moreover, the mean number of migrations has been analyzed regarding various amounts of communication with end-user devices. The rest of this paper is organized as follows. The motivation scenario is explained in Section II. In Section III, we discuss the related work and represent the system model Section IV. We formulate the ACM problem in optimization problem and also propose a heuristic in Section V and the experiment's results are reported in Section VI. Finally, we conclude the paper in Section VII.

## II. MOTIVATION SCENARIO

An earthquake early warning and recovery application which detects the earthquake in its early stage described in [2] is used as our motivation scenario. It is represented by the structured graph as shown in Fig 1.a. It is composed of 6 components including Early Warner, Data Analyzer, Warning Alert Issuer, Victim Detector, Victim Rescuer, and Historical Storage. The application can be described by a structured graph that defines which components are executed in parallel or sequence. For example, "Victim Rescuer" and "Historical Storage" are executed in sequence, whilst the composite of them are executed in parallel with "Victim Detector" and "Warning Alert Issuer". Readers are referred to [2] for the details. Fig.1.b and Fig. 1.c are examples of structured graphs of Flood warning and autonomous driving applications which are presented in[3] and[10]respectively.

Each one of these components can be hosted on either cloud or fog [2]. As fog nodes can be mobile (e.g., drones), we assume that drone 1 which hosts the "Early Warner & Analyzer" moves away from the end-users' devices (changing from Fig. 2.a to Fig. 2.b). If in the new location of drone 1, the distance between drone 1 and the end-user devices is d1 and we have drone 2 at a shorter distance d2, it might make sense to migrate the "Early Warner & Analyzer" to drone 2 to reduce latency. This is illustrated in Fig 2.b. This migration could be done by the Platform as a Service (PaaS) which provisions applications in the hybrid cloud/fog.

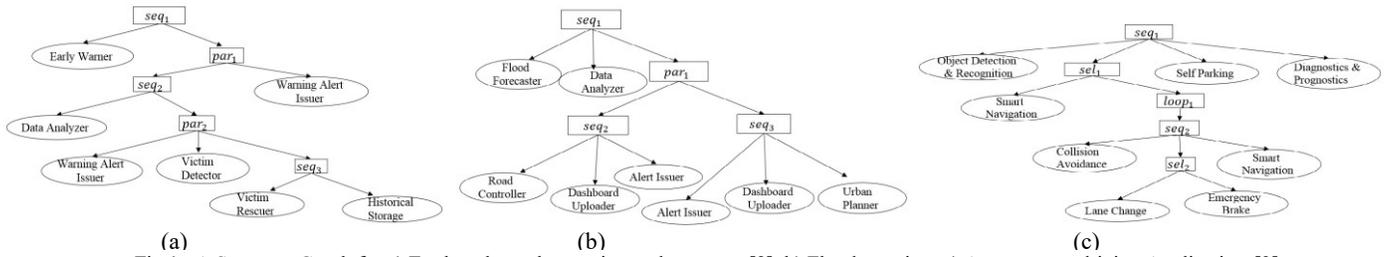

Fig.1. A Structure Graph for a) Earthquake early warning and recovery [2] b) Flood warning c) Autonomous driving Applications[3]

### III. RELATED WORK

In this section, we first review the relevant literature for application component migration in hybrid cloud/fog systems, and then we review the proposed solutions for the migration problem in other domains.

#### A. Migration in the Hybrid Cloud/Fog Domain

Very few works have considered the migration of application components in hybrid cloud/fog domains. Zhiqing *et al.* in [7] propose a migration solution to handle the mobility of end users while Zhu *et al.* in [5] do so by considering the mobility of fog nodes. In [7], Zhiqing *et al.* present a container migration algorithm based on reinforcement learning. Their goal is to reduce the migration cost, power consumption, and communication latency. Zhu *et al.* in [5] propose an algorithm to dynamically distribute application tasks across stationary fog nodes, mobile fog nodes, and cloud nodes. When a mobile fog node hosting a task moves, the task is migrated to another fog node. The goal is to minimize application latency and quality loss.

Both works consider migration in the context of mono-component applications called tasks. However, in the real world, most applications are composed of several components as shown by the previous scenario. The interactions between the components need to be taken into account as we do in our proposed solution.

#### B. Migration in Domains other than Cloud/Fog

There are several works that investigate the migration problem in domains other than the cloud/fog domain. The migration problem in the NFV domain has been studied widely over the last few years. Cho *et al.* in [11] propose a VNF migration algorithm to minimize the network latency in dynamic networks where the availability of resources changes rapidly. Xia, *et al.* in [12] model the VNF migration problem as an Integer Linear programming (ILP) and propose a heuristic to find an optimal migration plan. Their objective is to reduce the migration cost. However, none of these two works consider the interactions between the application components when making the migration decision. In contrast, in our work, we propose a migration solution that considers a set of interacting application components that can interact using multiple sub-structures (e.g., selection and loop).

The authors of [13] and [14] focus on the migration problem to handle the mobility of the end-users. They aim to provide a solution to the migration problem by considering edge computing. In [13], Rodrigues *et al.* propose an analytical model for minimizing the perceived service delay of end-users in an edge computing environment. This model performs Virtual Machine (VM) migration and adjusts the transmission power in order to improve the processing and the transmission delays. The authors of [14] study the service migration problem in the edge cloud in response to the user movement and network performance. The solution is based on Markov decision process. Although these solutions address the migration problem, they only consider stationary edge nodes with predefined locations. This assumption makes their approach nonfunctional when the system includes mobile nodes.

### IV. SYSTEM MODEL

In this section, we describe the modeling of applications, cloud, fog, and end-user devices.

**Applications** – We assume each application is composed of several components where each component is implemented as a VNF. Let $K$ be the set of VNF types such as victim detector, warning alarm issuer, etc. Each VNF type $f^k$ has a traffic processing capacity $c_{f^k}$. we define $\mu_{f^k}$ as the maximum allowable utilization of the capacity. $s_{f^k}$ is the size of VNF $f^k$.

The whole application is modeled as a chain of VNFs, i.e., VNF-FG. The cloud and the fog offer NFVIs that can host VNFs. Each VNF-FG is converted to a tree structure. In the tree structure, the leaf node represents the application component and the internal nodes represent one of the substructures; sequence, parallel, selection, or loop [4]. Let us consider $Req$ as a set of structured VNF-FG requests received by the system. The set of required VNF types for request $R$ is represented by $vnf_R \subset K$. The immediate predecessors of VNF $f^k$ is represented by $IP(f^k)$. VNF $f^k$ can be executed when all immediate predecessors have been executed and the required data from the immediate predecessors and end-user devices have been transmitted to the VNF host. For request $R$, the traffic rate from a predecessor VNF; $IP(f^k)$, to VNF $f^k$ is represented by $A^R_{ip(f^k),f^k}$.

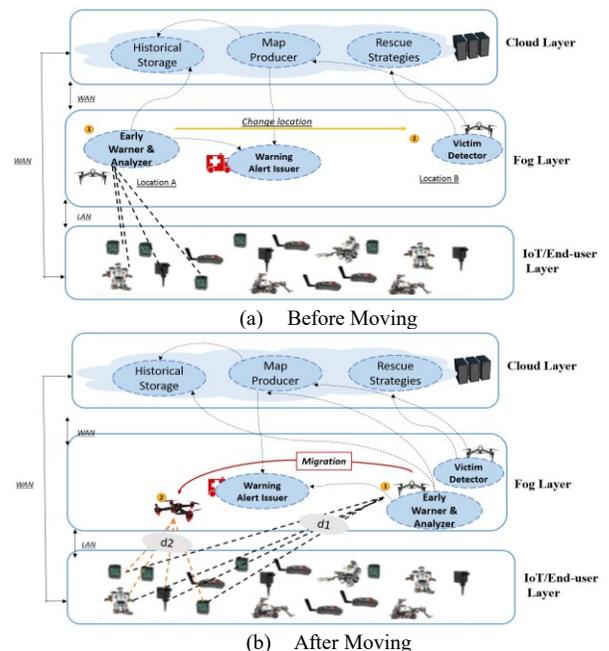

Fig. 2. Migration of one component

**Cloud/Fog** - Let us consider the cloud and fog as an undirected graph $G = (N^z, E^z)$. Here, $z = C|F$ can denote either cloud or fog. The set of cloud/fog nodes are $N^z$, and $E^z$ is the set of edges that shows the links between nodes. Each node has a capacity (memory, CPU, etc.) and a maximum allowed utilization of that capacity, represented by $c_{n^z}$ and $\mu_{n^z}$, respectively. $D^{f^k}_{n^z}$ is the delay of processing one unit of traffic for VNF type $k$ located on a cloud/fog node $n^z \in N^z$. $E^J$ is a set of all possible communications in the network including the edges between the cloud and the fog nodes. Let $n_l^z.l(\tau)$ be the location of $n_l^z$ at time slot $\tau$. The location is fixed for all cloud nodes during time slots. However, it varies for mobile fog nodes. The propagation delay and communication bandwidth between two nodes $n_l^z$ and $n_m^z$ at time slot $\tau$ depends on the location of the nodes and are assumed to be given by $PD_{e^J_{lm}}(n_l^z.l(\tau), n_m^z.l(\tau))$ and $BW_{e^J_{lm}}(n_l^z.l(\tau), n_m^z.l(\tau))$, respectively. Similarly, we define $D_{e^J_{lm}}(n_l^z.l(\tau), n_m^z.l(\tau))$ as the delay (including propagation and transmission) per unit of traffic transmission between the nodes $n_l^z$ and $n_m^z$.

**End-users' devices** – The set of end-users communicating with the components in request $R$ is represented as $U_R$. For simplicity, we assume that end-user devices are fixed. The variable $\omega^R_{u \times f^k} \in \{0,1\}$ is 1 if there is communication between $u$ and $f^k$ for request $R$. The traffic rate between end-user $u$ and the VNF $f^k$ of request $R$, is represented by $A^R_{u \times f^k}$. The bandwidth/delay of the mentioned communication at time slot $\tau$ depends on the location of the node hosting the VNF. The bandwidth and the delay are represented by $BW_{e^{u,n^z}}(n^z.l(\tau))$ and $D_{e^{u,n^z}}(n^z.l(\tau))$, respectively. Note that if the VNF is hosted on a cloud node, the bandwidth will be constant. Finally, we represent the maximum allowed link utilization for the mentioned communication by $\mu_{e^{u,n^z}}$.

## V. APPLICATION COMPONENT MIGRATION

In this section, we model the migration problem as an optimization problem with the objective of minimizing the aggregated makespan of the applications. The makespan for an application is the time between starting the execution of the first component until the end of the execution of the last component [4]. In the rest, we explain the makespan calculation and we define the objective function and the constraints. Finally, we present a heuristic to solve the optimization problem.

### A. Makespan Calculation

We assume that time is divided into time slots with the length of $|T|$. The binary variable $x_{i,f^k,n^z}(\tau)$ is 1 when instance $i$ of VNF $f^k$ is deployed on the cloud/fog node $n^z$ at time slot $\tau$; otherwise, it is 0. The binary variable $x^R_{i,f^k,n^z}(\tau)$ is 1 when the instance $i$ of VNF $f^k$ that has been deployed on cloud/fog node $n^z$ is assigned to request $R$ in time step $\tau$; otherwise, it is 0. To calculate the makespan of a request, VNFs processing time, VNFs/end-users' devices communication times, and migration time need to be calculated. The total incoming traffic to a VNF is the summation of traffic from its predecessors and traffic from end-user devices. Eq. (1) shows this calculation.

$$A^R_{f^k} = \sum_{ip(f^k) \in IP(f^k)} A^R_{ip(f^k),f^k} + \sum_{u \in U_R} A^R_{u \times f^k} \quad (1)$$

Let $\tau_p$ be the time slot at which the processing of VNF $f^k$ is computed. $\tau_p$ is the smallest integer that satisfies the constraint as shown in Eq. (2). Note that the traffic which can be processed in a single time slot is calculated by $\frac{|T|}{D^{f^k}_{n^z}}$ when the VNF is located on node $n_z$. In this regard, $\tau_p$ is the time slot at which the accumulated processed traffic is equal or greater than the input traffic to the VNF as defined in Eq. (2). In this regard, the time at which the processing of VNF $f^k$ of request $R$ is completed is calculated by Eq. (3).

$$\sum_{\tau=0}^{\tau_p} \sum_{n^z \in N^z} \sum_{i \in I_{f^k}} \frac{x^R_{i,f^k,n^z}(\tau)|T|}{D^{f^k}_{n^z}} \geq A^R_{f^k} \quad (2)$$

$$M_{proc}(R, f^k) = \tau_p \cdot |T| \quad (3)$$

The VNFs processing time for the whole application can be calculated by traversing the tree structure from the leaves to the root and aggregating the processing times. When a node in the tree is a substructure, namely $S_i$ (sequence, parallel, selection, or loop), the processing time for that node $M_{proc}(R, S_i)$ is calculated as Eq. (4). Here, $CH(S_i)$ are the children set of $S_i$ in the tree, $q$ is the probability of loop repetition, and $h_{S_j}$ is the probability of a child $S_j$ selection.

$$M_{proc}(R, S_i) = \begin{cases} \sum_{S_j \in CH(S_i)} M_{proc}(R, S_j) & S_i \text{ is seq} \\ \max_{S_j \in CH(S_i)} M_{proc}(R, S_j) & S_i \text{ is par} \\ \sum_{S_j \in CH(S_i)} h_{S_j} \cdot M_{proc}(R, S_j) & S_i \text{ is sel} \\ \frac{q}{1-q} \cdot \sum_{S_j \in CH(S_i)} M_{proc}(R, S_j) & S_i \text{ is loop} \end{cases} \quad (4)$$

In this regard, $M_{proc}(R, root)$ gives the makespan for request $R$.

The traffic transmission from immediate predecessors of a VNF $f^k$ i.e., $IP(f^k)$ to that VNF is completed at time slot $\tau_{cp}(ip(f^k))$ which is the smallest integer that satisfies constraint (5). Note that the traffic which can be transmitted in a single time slot is calculated by $\frac{|T|}{D_{e^J_{lm}}(n_l^z.l(\tau), n_m^z.l(\tau))}$ when the immediate predecessor and the VNF have been located at nodes $n^{Z_l}$ and $n^{Z_m}$, respectively Mathematically, the transmission is completed when the accumulated transmitted traffic is equal to or greater than the sent traffic.

$$\sum_{\tau=0}^{\tau_{cp}(ip(f^k))} \sum_{n^Z \in N^Z} \sum_{i \in I_{f^k}} \frac{x^R_{i,IP(f^k),n^{Z_l}}(\tau) \cdot x^R_{j,f^k,n^{Z_m}}(\tau)|T|}{D_{e^J_{lm}}(n^z_l.l(\tau), n^z_m.l(\tau))} \geq A^R_{ip(f^k),f^k} \quad (5)$$

Similarly, the transmission of traffic $A^R_{u \times f^k}$ from end-user $u$ to a VNF $f^k$ belonging to a VNF-FG request $R$, is completed at time slot $\tau_{cu}(u)$. The time slot is the smallest integer satisfying the bellow constraint:

$$\sum_{\tau=0}^{\tau_{cu}(u)} \sum_{n^Z \in N^Z} \sum_{i \in I_{f^k}} \frac{x^R_{i,f^k,n^z}(\tau)|T|}{D_{e_{u,n^z}}(n_z.l(\tau))} \geq A^R_{u \times f^k} \quad (6)$$

The total communication time of VNF $f^k$ belonging to request $R$ with the predecessor VNFs and end-users is calculated as below:

$$M_{com}(R, f^k) = \max\left(\max_{ip(f^k) \in IP(f^k)} \tau_{cp}(ip(f^k)), \max_{u \in U_R} \omega^R_{u \times f^k} \cdot \tau_{cu}(u)\right) \cdot |T| \quad (7)$$

In this regard, $M_{com}(R, root)$ is calculated by aggregating the communication time of the VNFs similar to what discussed about the processing time based on Eq. (4).

Although the containerization technology has reduced the migration overhead [9], we have included the migration time in makespan calculation in order to have a more precise analysis. The consumed time for migrating VNF $f^k$ belonging to request $R$ during application execution is calculated by Eq. (8).

Here, $\tau_{Max}$ is large enough to cover the execution of all requests. The value can be achieved through statistics. When the VNF is migrated from node $n^z_l$ to $n^z_m$ in time slot $\tau$, the propagation delay and VNF transmission delay is involved in migration time calculation.

$$M_{mig}(R, f^k) = \sum_{\tau=0}^{\tau_{Max}} \sum_{n^z_l, n^z_m \in N^Z, m \neq l} x^R_{i,f^k,n^z_l}(\tau) \cdot x^R_{i,f^k,n^z_m}(\tau+1) \cdot [PD_{e^J_{lm}}(n^z_l.l(\tau), n^z_m.l(\tau)) + \frac{S_{f^k}}{BW_{e^J_{lm}}(n^z_l.l(\tau), n^z_m.l(\tau))}] \quad (8)$$

We should consider that $M_{mig}(R, root)$ is calculated by aggregating the migration time of the VNFs similar to processing and communication time calculations. In this regard, the makespan of request $R$ is calculated as below:

$$M(R) = M_{proc}(R, root) + M_{com}(R, root) + M_{mig}(R, root) \quad (9)$$

### B. Objective Function and Constraints

The objective function is defined as the summation of the makespan of all requests and migration overhead as below:

$$Min \sum_{R \in Req} M(R) \quad (10)$$

Equations (2), (5), and (6) are the constraints. Furthermore, constraint (11) ensures that the total resources required by instances of all VNF types deployed on a cloud/fog node do not exceed the capacity of the hosting node.

$$\sum_{k \in C} \sum_{i \in I_{f^k}} \vartheta_{f^k} \cdot x_{i,f^k,n^z}(\tau) \leq \mu_{n^z} \cdot c_{n^z} \quad \forall n^Z \in N^Z, \tau \quad (11)$$

In constraint (12), we ensure that the chain traffic transmitted through the network does not exceed the link utilization limit. A similar constraint exists for the communication between end-users and cloud/fog nodes. i.e., Eq. (13).

$$\sum_{R \in Req} A^R_{ip(f^k),f^k} \cdot x^R_{i,ip(f^k),n^{Z_l}}(\tau) \cdot x^R_{j,f^k,n^{Z_m}}(\tau) \leq \mu_{e^J_{lm}} \cdot BW_{e^J_{lm}}(n^{Z_l}.l(\tau), \forall e^J_{lm} \in E^J, \tau \quad (12)$$

$$\sum_{R \in Req} A^R_{u \times f^k} \cdot x^R_{i,f^k,n^z}(\tau) \cdot \omega^R_{u \times f^k} \leq \mu_{e_{u,n^z}} \cdot BW_{e_{u,n^z}}(n_z.l(\tau)) \forall u \in U_R, n^Z \in N^Z, \tau \quad (13)$$

Eq. (14) ensures that the capacity of an instance of a VNF $f^k$ is not overloaded by the total traffic received from immediate predecessor(s) and end-user devices total requests.

$$\sum_{\forall R \in Req} A^R_{f^k} \leq \mu_{f^k} \cdot c_{f^k} \quad \forall k \in K, \forall n^Z \in N^Z, \tau \quad (14)$$

Eq. (15) ensures that the assigned VNF instances are already deployed in the network,

$$x^R_{i,f^k,n^z}(\tau) \leq x_{i,f^k,n^z}(\tau)$$
$$\forall R \in Req, f^k \in vnf_R, i \in I_{f^k}, n^Z \in N^Z, \tau \quad (15)$$

### C. A Heuristic for Component Migration

As the solution space of the problem is exponential, a heuristic is required to solve the problem in an efficient way. In this paper, we propose a heuristic which is performed in an iterative manner. Each iteration performs migration for one-time slot and it starts with an initial placement (initial values for variables $x^R_{i,f^k,n^z}(\tau)$ and $x_{i,f^k,n^z}(\tau)$). In each iteration, the values of the mentioned variables are changed according to the new fog nodes' location. For the current time slot, the same values as the previous time slot are assigned to the variables. Then, for each request, the VNFs whose interactions with their immediate predecessors/end-users have not been completed by the current time slot are considered. We look for the fog nodes with enough capacity to process the VNF traffic and receive traffic from predecessor VNFs. Among those candidates, a fog node is selected which has the minimum communication time with predecessors/end-users and VNF processing, as well as migration time from the current VNF host to that node. If the selected node does not already have a VNF of that type, a VNF type at that node is instantiated and the VNF will be assigned to that instantiation. If the selected node already has some instances of the VNF type, a random instance with enough capacity is selected. In the case that the previous host does not serve any request of that VNF type, the variable indicating the instantiation of that VNF in the previous host gets the value of 0. Fig.3 illustrates the pseudo code.

I. PERFORMANCE EVALUATION

### A. Simulation Setup

The network in the simulation consists of 4 end-user devices, 2 cloud nodes, and 3 fog nodes. We assume that every two nodes

```
Input:    time-slot τ ; x^R_{i,f^k,n^z}(τ-1) {Current allocation}; x_{i,f^k,n^z}(τ-1)
{Current VNF instantiation} ; n^z. l(τ)
Output: x^R_{i,f^k,n^z}(τ) {Next allocation }; x_{i,f^k,n^z}(τ) { Next VNF instantiation}
Procedure:
x^R_{i,f^k,n^z}(τ) ← x^R_{i,f^k,n^z}(τ-1)   ∀R,i,…
For each request R do
  For each VNF f_k belonging to request r do
    If the interaction of f_k with End-users' devices/predecessors have not been
completed by time step τ
      n_0^f ← argmax_{n^z} x^R_{i,f^k,n^z}(τ-1) // Current host of VNF
      i ← the associated instance
      CandidateSet ← Fog nodes with enough capacity to process f_k traffic
and receive traffic from immediate predecessor VNFs/End user device
      n_1^f ← Find a node in CandidateSet which minimizes the summation of
End user device/predecessors communication latency, VNF processing delay
and migration overhead from n_0^f to n_1^f
      If n_1^f has not already have a VNF type f_k
        Create a new instance of f_k
        j ← index of the instance
        x_{j,f^k,n_1^f}(τ) ← 1
      End-If
      If n_1^f has already have a VNF type f_k
        j ← Select a random instance with enough capacity
      End-If
      Move f_k to n_1^f i.e., x^R_{i,f^k,n_0^f}(τ) = 0; x^R_{j,f^k,n_1^f}(τ) = 1
      If n_0^f does not host for any other request which needs VNF f_k
        x^R_{j,f^k,n_0^f}(τ) ← 0
    End-If
  End-For
End-For
```

Fig. 3. Heuristic pseudocode

in the network can directly communicate with each other (a fully connected graph topology). We have set the network specification similar to the literature [2],[3],[4],[15]. All fog/cloud nodes and end-user devices have been assumed to be deployed in a square environment. The time slot duration is chosen to be 0.05 ms. At each time slot, each fog node is located in a random location in the deployment area.

Each cloud node has 8 VCPUs and each fog node has 3VCPUs. The cloud nodes communicate with each other with the bandwidth of 100Gbps. The fog nodes utilize a bandwidth of 100Mbps to communicate with each other. Finally, the bandwidth between cloud and fog nodes is 10Gbps. The propagation delay in the network has been chosen in the range of 0.1 to 0.6 msec such that the nodes with more distance from each other have more propagation delay for communication. The processing delay per unit of traffic (1 KB/sec) in cloud and fog nodes has been set to 3.12 sec and 0.03 sec, respectively.
We have performed the simulation for 3 real applications [4]: 1) earthquake early warning application with 6 VNFs; 2) Flood warning application with 6 VNFs; 3) Autonomous driving application with 7 VNFs (see Fig. 1). Equal probability has been assumed for selection substructure. VNF instances require a random number of VCPUs between 1 and 4. Similar to [16], the VNF image size is assumed to be 13 MB. The traffic load of end-user devices toward connected VNFs to the end-user devices is 80KB/sec. The end-user devices communicate with fog and cloud with 54Mbps and 10Gbps bandwidth capacity respectively. we consider the probability of collision risk $q=$

Table I. Comparing the makespans of the Heuristic, Random Migration, and No-Migration and comparing sum of migration with the proposed heuristic and Random Migration

| Apps | Makespan(ms) | | | Mean Number of Migrations Per Time Slot | |
|------|------|--------|------|------|--------|
|      | ACM  | Rand-M | No-M | ACM  | Rand-M |
| App1 | 4.42 | 14.59  | 6.35 | 0.43 | 0.5    |
| App2 | 3.84 | 13.47  | 4.81 | 0.38 | 0.48   |
| App3 | 3.96 | 11.03  | 5.13 | 0.46 | 0.74   |

0.25, $it$=0.33 and we give equal probabilities for selection; $p$=0.5.

### B. Evaluation Results

We compare our proposed approach for Application Component Migration (ACM) with two other approaches: "No-M", and "Rand-M". In "No-M", the initial placement is kept within the time slots. Note that we have randomly placed the components at nodes in the first time slot. In "Rand-M", the components are migrated randomly between cloud/fog nodes during time slots.

Table I Shows the makespan and the mean number of migrations per time slot for each application. Note that the first and last VNF of each application are connected to 4 end-user devices. As we can see, Rand-M has the lowest performance. The reason is that it randomly migrates the VNF instances between cloud/fog nodes without considering the effect on the makespan. On the other hand, the ACM achieves the lowest makespan. The reason is that it migrates the components according to the relative geographical locations of fog nodes to end-user devices, as well as cloud nodes which reduces the communication time and the makespan, consequently. The number of migrations in the ACM is less than Rand-M since it avoids extra migrations that increase the makespan. Note that the values below 1 indicate that in some time slots no migration has happened.

Fig. 4 shows the impact of end-user devices' connections in the makespan for App1. We have increased the number of connections between the first/last VNFs and the end-user devices from 1 to 15. As expected, in all methods the makespan increases as well. However, the slope of such increment in "ACM" is significantly less than that of the other methods. This illustrates the effectiveness of the migration of components as advocated by the proposed method especially for a high volume of connectivity with end-users. Fig. 5 shows the makespan for App2. We have assumed that in the case of connectivity, each VNF is connected to 4 end-user devices. Then, we changed the number of connected VNFs to end-user devices from 2 to 6. When more VNFs are connected to end-user devices, the latency of communication with devices increases the makespan in all methods. The performance of ACM to other methods increase with more connected VNFs.

Fig. 6 shows the mean number of migrations per time slot in ACM for the three mentioned applications. We have changed the number of connected VNFs to end-user devices from 2 to 6. As the number of connections with end-user devices increases, more VNF instances will probably become farther from the end-user devices during application execution. This happens as a result of the mobility of their hosts in the fog domain. Therefore, the number of migrations increases as well to move the VNF instances closer to fog nodes and reduce the makespan.

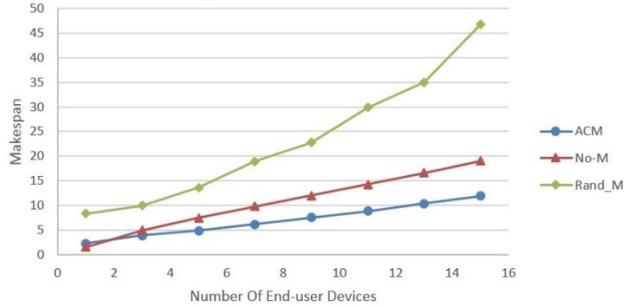

Fig. 4. Comparison of Makespan of three methods for Earthquake early warning and recovery application

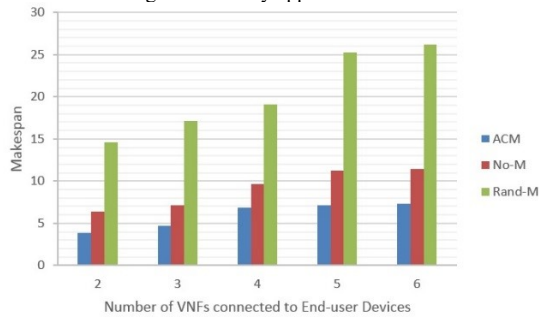

Fig. 5. Makespan when various number of VNFs in Flood warning application are connected to end-user devices

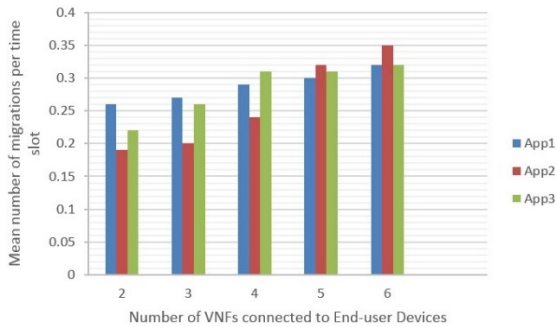

Fig. 6. Mean number of migrations per time slot in ACM

## II. Conclusion

In this paper, we have addressed the problem of application component migration in NFV-based hybrid cloud/fog system. Components are implemented as VNFs. Regarding the geographical location of fog nodes, we have calculated VNFs processing time, the communication time and migration time. To calculate the makespan, we have aggregated the processing communication and migration times by traversing a structured tree built over the input VNF-FG. We have modeled the problem as an optimization problem which aims at minimizing the makespan. We have proposed a heuristic to solve the problem. Simulation results show that the proposed migration method improves the makespan in comparison with Random-Migration and No-Migration approaches. Furthermore, we have analyzed the effect of communicating with end-user devices on the number of migrations. In the future, we plan to extend the framework so that other criteria, such as resource consumption cost, are also considered in the migration decision.

### Acknowledgment

This work is partially funded by the CISCO CRC program (Grant #973107), the Canada Research Chair Program, and the Canadian Natural Sciences and Engineering Research Council (NSERC) through the Discovery Grant program.